\documentclass{article}

\usepackage[T1]{fontenc}
\usepackage{times}

\usepackage{subfigure}
\usepackage[colorlinks=true, linkcolor=blue, citecolor=cyan, urlcolor=blue]{hyperref}
\usepackage{cite}
\usepackage{amsmath}
\usepackage{geometry}
\usepackage{graphicx}
\usepackage{float}
\usepackage{indentfirst}
\title{Insights into Neutron Star Matter: EoS Models and  Observations}
\author{Zuhua Ji, Jiarui Chen, Gaojian Wu\\
	Department of physics, School of Physical and Mathematical Sciences, \\Nanjing Tech University, Nanjing 210009, PR China}
\date{}
\begin{document}
	\maketitle
	
	\begin{abstract}
		The equation of state (EoS) for neutron stars is a crucial topic in astrophysics, nuclear physics, and quantum chromodynamics (QCD), influencing their structure, stability, and observable properties. This review classifies EoS models into hadronic matter, hybrid, and quark matter models, analyzing their assumptions, predictions, and constraints. While hadronic models characterize nucleonic matter, potentially including contributions from hyperons or mesons, hybrid models introduce phase transitions to quark matter, and quark models hypothesize the presence of deconfined quark matter cores or entirely quark-composed stars. By synthesizing results from recent theoretical and observational studies, this review aims to offer a comprehensive understanding of the methodologies used in constructing neutron star EoS, their implications, and future directions.
	\end{abstract}

	\section{Introduction}
	Neutron stars serve as natural laboratories for studying ultra-dense matter, providing insights into the behavior of matter under extreme conditions that cannot be replicated on Earth\cite{ref1}. These objects are the remnants of massive stellar explosions and contain densities exceeding nuclear saturation density, making their internal structure and composition a subject of significant interest in astrophysics and nuclear physics\cite{ref2}. Figure 1 adapted form \cite{ref14} shows the typical neutron star structure of different types at 1.4 and 2 solar masses.
	
	\begin{figure}[H]
		\centering
		\includegraphics[width=0.65\textwidth]{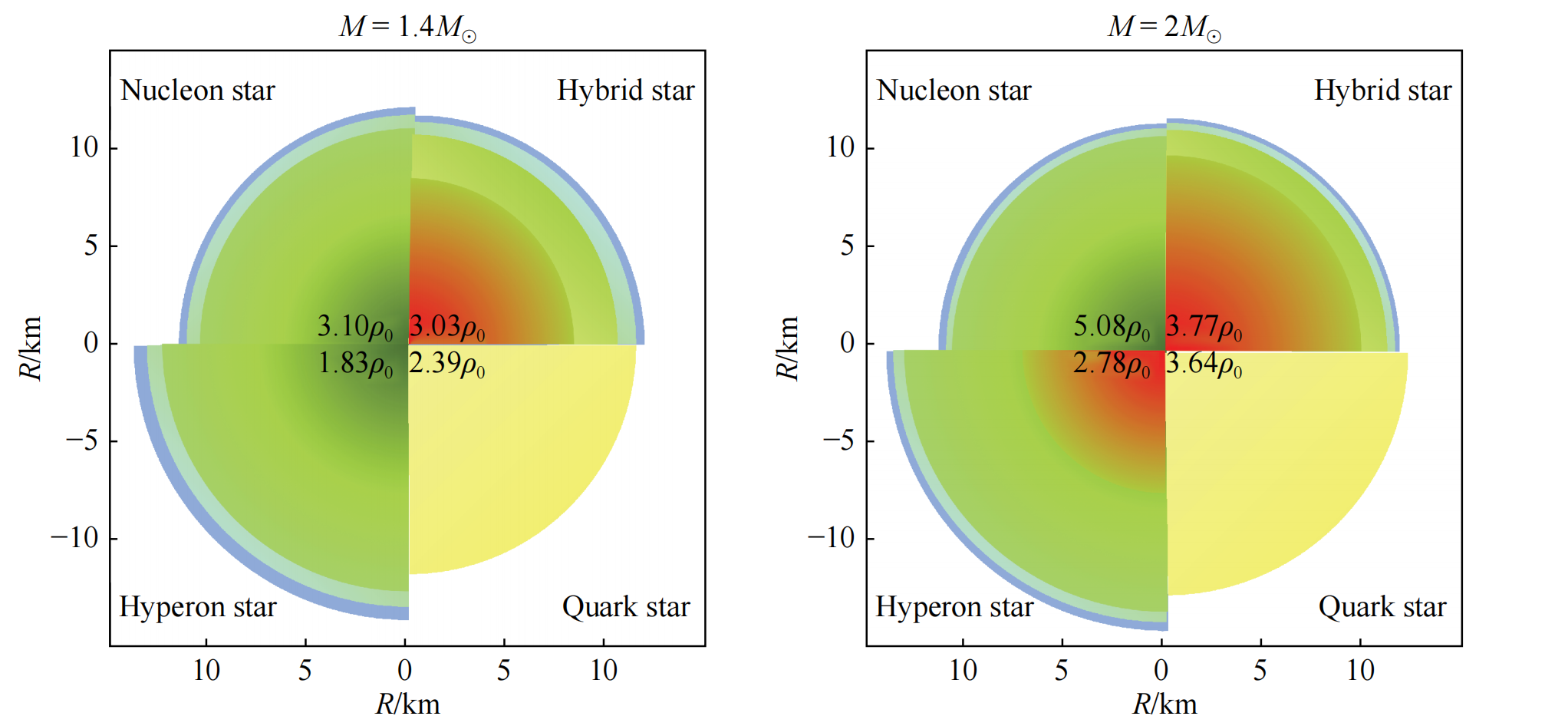}
		\caption{Typical neutron star structure at 1.4 and 2 solar masses}
		\label{fig1}
	\end{figure}
	
	Understanding the equation of state (EoS) of neutron stars is crucial for determining key properties such as their mass-radius relation, maximum mass, and tidal deformability, all of which have direct implications for both fundamental physics and astrophysical observations. Over the decades, numerous theoretical models have been proposed to describe the internal composition of neutron stars. Since the pioneering work of Oppenheimer and Volkoff in 1939\cite{ref3}, who introduced the first relativistic EOS based on a free neutron Fermi gas—thereby laying the groundwork for the Tolman–Oppenheimer–Volkoff (TOV) equations—there has been a continuous evolution in modeling dense matter. In 1998, Akmal et al.\cite{ref4,ref5} developed a variational approach to nuclear interactions, resulting in the widely adopted APR model, which remains a benchmark for nucleonic EOSs. Building on this, Read et al.\cite{ref20} in 2009 constructed the ALF1–ALF4 hybrid models by combining APR nuclear matter with color-flavor-locked quark matter, offering structured EOSs suitable for Bayesian inference.
	
	Efforts to improve precision and realism have continued. Gandolfi et al.\cite{ref5} applied auxiliary field diffusion Monte Carlo (AFDMC) methods in 2014 to calculate the neutron matter EOS with high accuracy, emphasizing the significance of three-body forces and symmetry energy. In 2017, Kaltenborn et al.\cite{ref16} introduced the DD2F-SF family, which incorporates a first-order hadron–quark phase transition via the string-flip model. Holographic techniques entered the scene in 2020, when Ecker et al.\cite{ref18} proposed hybrid EOSs based on the V-QCD framework, marking one of the first applications of the AdS/CFT correspondence to dense matter physics. That same year, Otto et al.\cite{ref9} adopted a functional renormalization group (FRG) approach within quark-meson models, enabling the inclusion of quantum and density fluctuations beyond the mean-field level.
	
	More recently, Kojo et al.\cite{ref8} presented the QHC21 EOS in 2022, advocating a quark–hadron crossover scenario that accounts for NICER's observation of nearly constant neutron star radii without requiring strong first-order phase transitions. Continuing this trend toward observationally guided modeling, Pang et al.\cite{ref15} in 2023 developed the NMMA framework, which integrates gravitational wave, kilonova, and X-ray data to constrain EOSs in a data-driven manner. These EOSs can present series of relationships like what is shown in figure 2 from Ref.\cite{ref1}, which illustrates the radius–mass relationship of some models.
	
	\begin{figure}[H]
		\centering
		\includegraphics[width=0.5\textwidth]{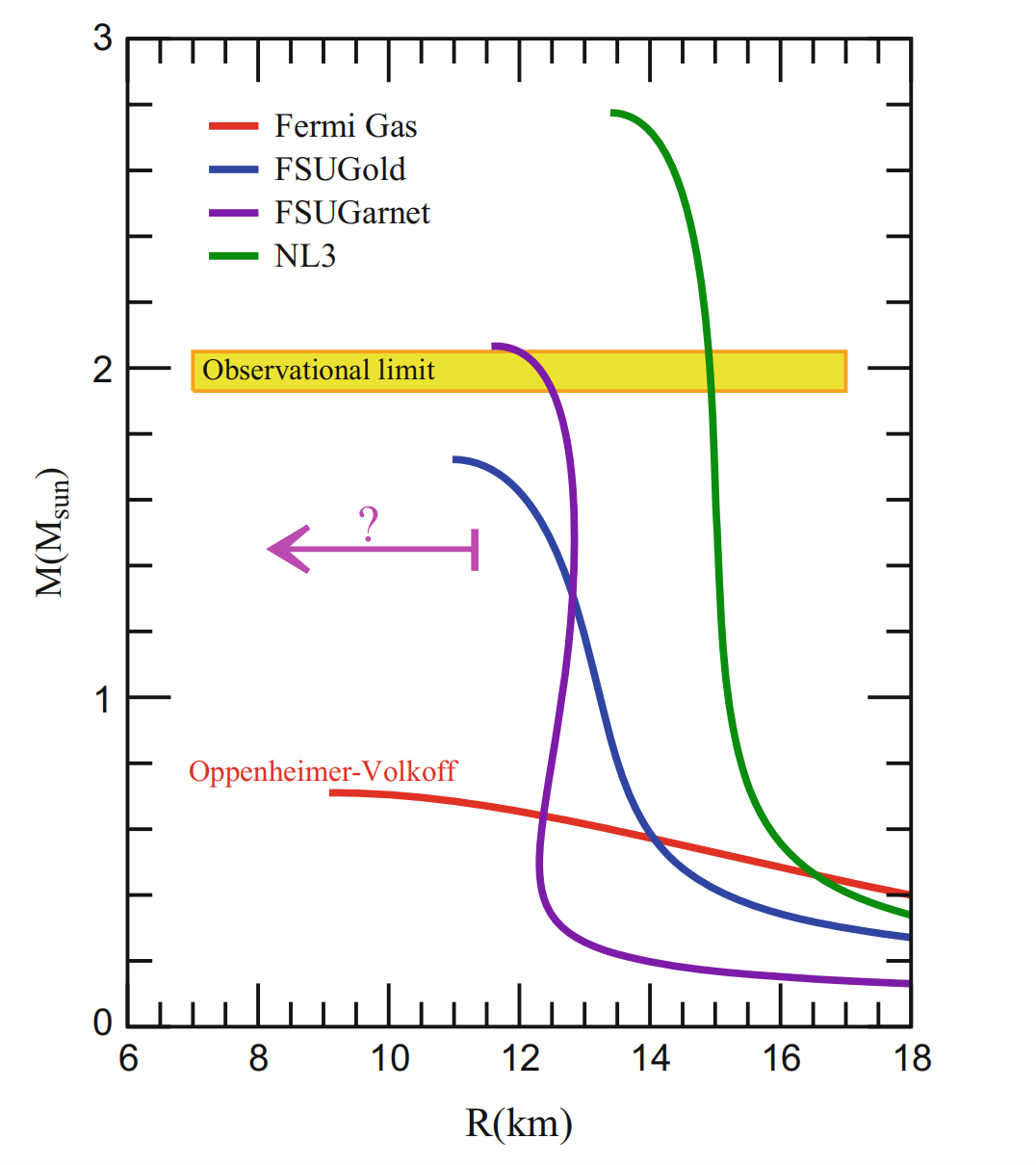}
		\caption{Radius–mass relationship of neutron stars shown in some models}
		\label{fig2}
	\end{figure}
	
	Recent advancements in observational techniques, including gravitational wave detections from binary neutron star mergers (e.g., GW170817)\cite{ref19} and precise X-ray timing measurements of radius and mass from NICER(e.g., J0740+6620)\cite{ref18}, have provided unprecedented constraints on the neutron star EoS. These observations, combined with theoretical developments in nuclear physics and quantum chromodynamics (QCD), have led to significant progress in understanding the microphysics governing neutron stars.
	
	This review aims to offer a comprehensive understanding of the methodologies used in constructing neutron star EoS, their implications, and future directions. Sec. \ref{2} presents a classification of the models by type and summarizes their key results, which also provides a comparative overview of these results. Then in Sec. \ref{3} we describe the constraints imposed on these models by astronomical observations and terrestrial high-energy experiments. Sec. \ref{4} outlines future observational approaches and potential directions for model development. Conclusions are drawn in Sec. \ref{5}.
	
	\section{Neutron Star EoS Models}
	\label{2}
	\subsection{Hadronic Matter Models}
	Hadronic matter models describe neutron star interiors as being composed primarily of nucleons (neutrons and protons), with additional hadronic degrees of freedom potentially appearing at higher densities. These models rely on nuclear interactions, which are typically mediated by mesons, to define the equation of state (EoS) of neutron-rich matter under extreme conditions. The primary goal of hadronic EoS models is to describe how matter behaves at densities significantly exceeding those found in atomic nuclei while maintaining consistency with both terrestrial nuclear physics experiments and astrophysical observations.
	
	\subsubsection{Models based on the Relativistic Mean-Field (RMF) Approach}
	
	One of the foundational approaches in modeling hadronic matter is the Relativistic Mean-Field (RMF) Theory\cite{ref6}, which provides a self-consistent description of nuclear interactions through the exchange of mesons such as scalar ($\sigma$), vector ($\omega$), and isovector ($\rho$) mesons. Models based on the RMF method (Hereinafter referred to as RMF models.) incorporate the effects of relativity to accurately describe high-density matter, making them particularly useful for neutron star modeling. The basic assumptions of this model are as follows. First, nucleon interactions are mediated by meson exchange fields, particularly the scalar ($\sigma$) and vector ($\omega$, $\rho$) mesons. Then, a self-consistent field approximation is employed, in which nucleons interact through mean-field potentials rather than direct two-body interactions. Thirdly, density-dependent coupling constants are introduced to better reproduce nuclear saturation properties. Additionally, beta equilibrium and charge neutrality are assumed to determine the composition of neutron star matter.
	
	Different models of this type, such as GM1, DD2, DD2F and TM1, have been developed to fit experimental nuclear matter properties and to predict neutron star characteristics\cite{ref6,ref16}. RMF models generally predict a stiff EoS, meaning that they provide sufficient pressure to counteract gravitational collapse, thereby supporting neutron stars with larger masses and radii. However, despite their success, uncertainties remain regarding the treatment of meson-nucleon couplings, density-dependent interactions, and the role of higher-order corrections in extreme conditions. Figure 3 adapted from Ref.\cite{ref16} illustrates the differences among DD2F models with various parameters. The SF parameters of seven EOSs correspond to different onset and final densities of the first-order phase transition. The SF parameter represents different surface tension values in the model, ranging from 0 to $\infty$
	.
	
	\begin{figure}[H]
		\centering
		\subfigure[Radius-mass]{
			\includegraphics[width=0.45\textwidth]{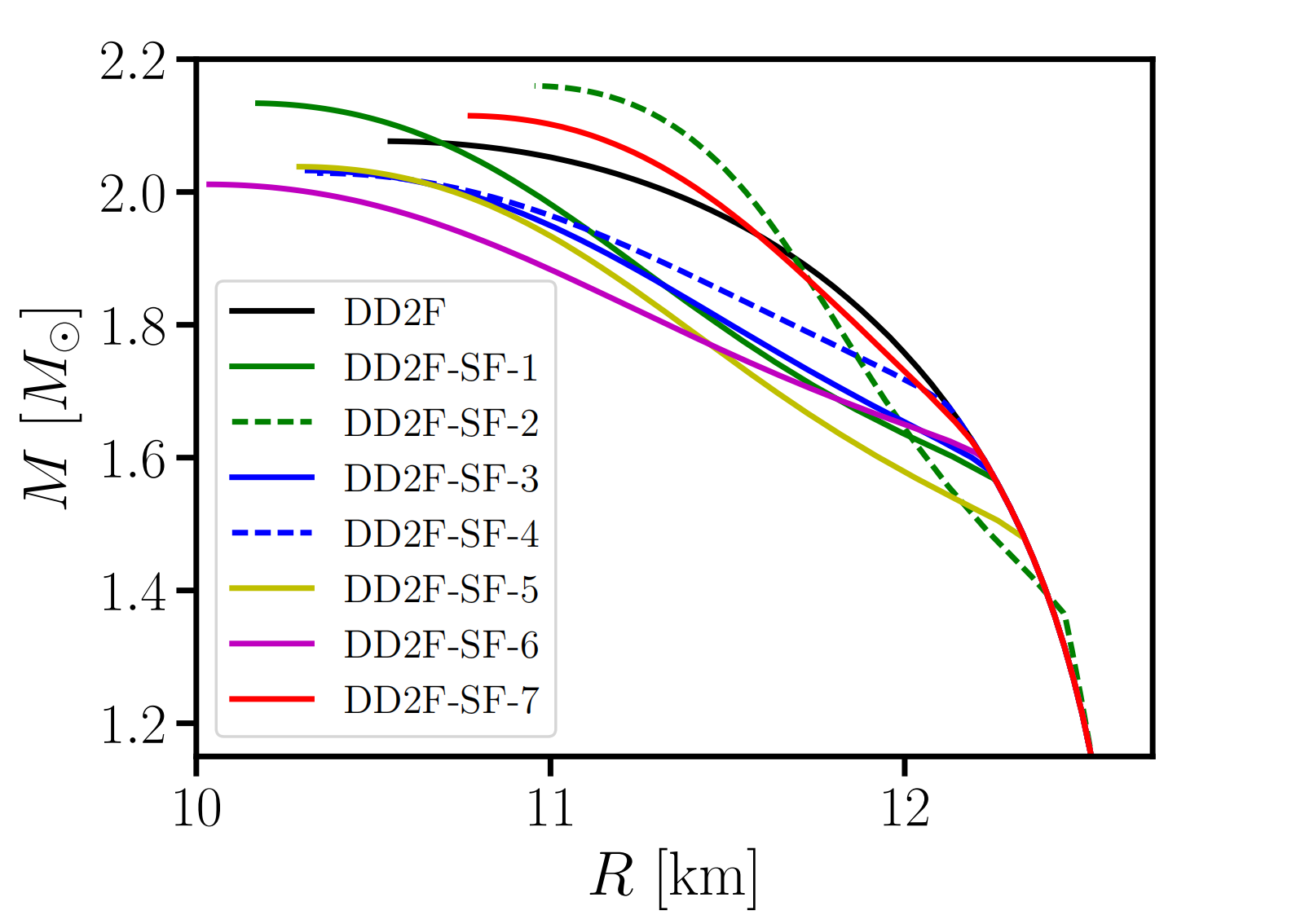}
		}
		\hspace{0.0\textwidth}
		\subfigure[Density-pressure]{
			\includegraphics[width=0.48\textwidth]{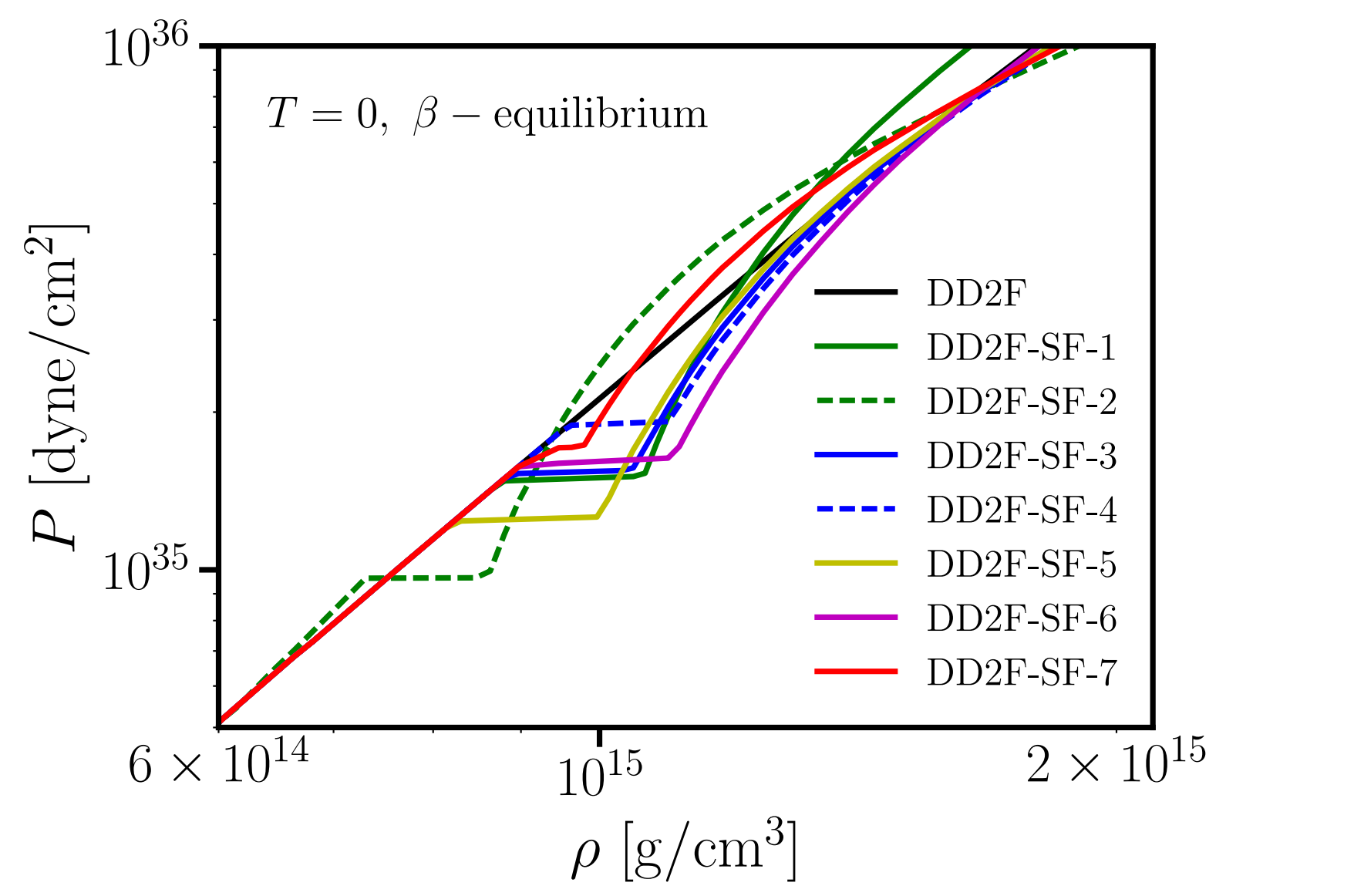}
		}
		\caption{(a) Radius–mass relationship and (b) Density-pressure relationship at different locations of the DD2F model under different parameters, \( M \) is the gravitational mass, \( R \) the circumferential eigen radius for nonrotating cold NSs. Solid green (black) curve displays the \( M \)-\( R \) relation for DD2F-SF-1(DD2F). Pressure as function of the rest-mass density for different EOSs of the DD2F-SF class. Black curve displays the purely hadronic reference model DD2F.}
		\label{fig3}
	\end{figure}
	
	Beyond nucleonic matter, RMF models also predict the possible presence of hyperons and meson condensates in neutron star interiors at sufficiently high densities. Hyperons ($\Lambda, \Sigma, \Xi$ baryons) become energetically favorable under extreme conditions, leading to a softening of the EoS and reducing the maximum mass a neutron star can support. This effect, known as the hyperon puzzle, poses a major challenge for hadronic models, as observed neutron stars with masses exceeding $2 M_{\odot}$ suggest that an overly soft EoS is unrealistic. To resolve this issue, researchers have introduced repulsive hyperon-nucleon interactions and additional vector meson couplings to counteract the softening effect while maintaining consistency with high-mass neutron star observations\cite{ref25}.
	
	Another exotic feature that RMF models consider is meson condensation\cite{ref11}, particularly pion ($\pi$) and kaon ($K$) condensation, which can occur at extremely high densities. The onset of meson condensation modifies the EoS by reducing the pressure at given energy densities, leading to additional phase transitions that impact neutron star cooling, thermal evolution, and structural stability. Kaon condensation, in particular, has been suggested as a potential mechanism for rapid neutron star cooling\cite{ref3} via enhanced neutrino emission, making it an important factor in astrophysical modeling.
	
	\subsubsection{Non-Relativistic Potential Models}
	
	While RMF models incorporate relativistic effects, an alternative approach to modeling hadronic matter is the Non-Relativistic Potential Models. These models rely on phenomenological nuclear interactions, such as the Skyrme force, or more microscopic approaches like variational methods and Brueckner-Hartree-Fock (BHF) calculations\cite{ref24}. Unlike RMF models, which use meson fields to mediate nuclear interactions, non-relativistic potential models describe nucleon interactions through direct empirical fits to experimental nuclear data. The basic assumptions of this model are as follows. First, two-body nuclear forces are dominant in determining neutron star EoS, often supplemented by phenomenological three-body forces. Then, effective density-dependent potentials are introduced to match nuclear experimental data. Thirdly, non-relativistic kinetic energy terms are used, making them suitable at lower densities but requiring extensions for ultra-dense matter. In addition, many-body effects such as three-nucleon interactions (TNI) play a crucial role in determining the stiffness of the EoS.
	
	The Skyrme models utilize effective density-dependent interactions that are calibrated using empirical nuclear matter properties. These models have been successful in describing the bulk properties of nuclear matter, but their applicability to extremely dense matter, such as that found in neutron star cores, remains limited due to uncertainties in extrapolating their interaction terms beyond nuclear saturation density.
	
	The Brueckner-Hartree-Fock (BHF) method is a more microscopic approach that directly incorporates two- and three-body nuclear forces derived from nucleon scattering experiments. By solving the Bethe-Goldstone equation, BHF calculations provide a more fundamental description of nuclear interactions, avoiding some of the empirical uncertainties present in Skyrme-type models. However, BHF calculations often predict a relatively soft EoS, leading to neutron star maximum masses that struggle to exceed the $2 M_{\odot}$ observational limit.
	
	Another widely used technique in non-relativistic modeling is variational methods, which attempt to determine the ground-state energy of neutron-rich matter by minimizing the total energy per particle. These methods use realistic nucleon-nucleon interactions, such as the Argonne V18 potential, in combination with phenomenological three-body forces to refine neutron star EoS predictions\cite{ref10}. Variational approaches have been particularly useful for understanding nuclear saturation properties, but they face challenges in describing extremely high-density matter where relativistic effects become significant.
	
	Non-relativistic potential models continue to be refined using advancements in nuclear many-body theory and new experimental constraints from heavy-ion collisions. However, their limitations in handling ultra-dense matter and their inherent non-relativistic nature make them less favored compared to RMF models when modeling the cores of neutron stars. Future improvements in these models may involve better constraints on three-body interactions and extensions that incorporate relativistic corrections to improve their high-density predictions.
	
	\subsubsection{Key Conclusions}
	
	Hadronic matter models assume that neutron star interiors are composed primarily of strongly interacting nucleons. These models generally produce relatively stiff equations of state, allowing neutron stars to support masses above \(2 M_{\odot}\), as required by pulsar observations. However, the inclusion of additional hadronic degrees of freedom, such as hyperons or meson condensates, tends to soften the EoS, reducing the maximum mass that a neutron star can support.
	
	In relativistic mean-field (RMF) models\cite{ref6}, nucleon interactions are mediated by meson exchange, and the energy density and pressure are determined by the self-consistent solution of the mean-field equations. The equation governing pressure in RMF models is given by:
	
	\begin{equation}
		P = \sum_i P_i + \frac{1}{2} m_{\sigma}^2 \sigma^2 - \frac{1}{2} m_{\omega}^2 \omega^2 - \frac{1}{2} m_{\rho}^2 \rho^2
	\end{equation}
	
	where the first term represents the contributions from baryons, and the remaining terms account for the interactions mediated by the scalar (\(\sigma\)), vector (\(\omega\)), and isovector (\(\rho\)) mesons. RMF models tend to predict neutron stars with radii in the range of 12–13 km for a \(1.4 M_{\odot}\) star, consistent with NICER observations, and they generally satisfy tidal deformability constraints from GW170817.
	
	The presence of hyperons (\(\Lambda, \Sigma, \Xi\)) in neutron star cores further complicates hadronic EoS predictions. The equilibrium condition for hyperons is given by:
	
	\begin{equation}
		\mu_B = \mu_n + \mu_e
	\end{equation}
	
	where \(\mu_B\) is the hyperon chemical potential, \(\mu_n\) is the neutron chemical potential, and \(\mu_e\) is the electron chemical potential. The introduction of hyperons softens the EoS, reducing the maximum neutron star mass to below \(2 M_{\odot}\), which conflicts with observations of high-mass pulsars. Additional repulsive interactions, such as hyperon-nucleon vector meson couplings, are often introduced to counteract this effect.
	
	Another feature in hadronic models is meson condensation, particularly pion (\(\pi\)) and kaon (\(K\)) condensation, which occurs at extremely high densities. The onset of kaon condensation is determined by:
	
	\begin{equation}
		\mu_K = \mu_e
	\end{equation}
	
	This process can reduce the pressure at a given energy density, leading to additional phase transitions that influence neutron star cooling and structural stability.
	
	\subsection{Quark Matter Models}
	
	Quark matter models describe neutron star interiors as being composed entirely of deconfined quarks rather than nucleons or hadrons\cite{ref7}. These models are based on quantum chromodynamics (QCD), which predicts that at sufficiently high densities, hadrons dissolve into their constituent quarks. The study of quark matter is crucial for understanding the possible existence of strange quark stars and color-superconducting phases, both of which have unique implications for neutron star observations and the equation of state (EoS).
	
	While purely quark matter stars have yet to be observed, theoretical and computational advances suggest that deconfined quark matter could exist in the cores of neutron stars. In this section, we introduce three major approaches to modeling quark matter in neutron stars: Nambu–Jona-Lasinio (NJL) models, Functional Renormalization Group (FRG) approaches, and holographic QCD models (V-QCD models).
	
	\subsubsection{Nambu–Jona-Lasinio (NJL) Model}
	
	The Nambu–Jona-Lasinio (NJL) model is one of the most widely used frameworks for describing quark matter. This model is based on an effective field theory approach to QCD, in which quark interactions are governed by chiral symmetry breaking and restoration. The NJL model captures key features of quark dynamics, such as the emergence of a constituent quark mass due to spontaneous chiral symmetry breaking, which significantly affects the stiffness of the EoS. The basic assumptions of this model are as follows. Firtst, chiral symmetry breaking and restoration govern the behavior of quark matter at high densities.Then, quark interactions are non-perturbative, described using four-fermion interactions. Thirdly, no confinement mechanism is included, meaning quarks are always deconfined in the model. Additionally, vector repulsion terms can be added to stiffen the EoS and increase maximum neutron star mass predictions.
	
	One of the major predictions of NJL-based EoS is that quark matter tends to be softer than hadronic matter, meaning that it provides less pressure support against gravitational collapse. This often results in difficulties in explaining the existence of neutron stars with masses exceeding \(2 M_{\odot}\). To address this issue, researchers introduce additional interactions, such as vector repulsion terms and diquark pairing effects, to stiffen the equation of state at high densities.
	
	Despite its usefulness, the NJL model has some limitations. It lacks confinement—a fundamental aspect of QCD—meaning that it cannot fully capture the transition from hadronic matter to deconfined quark matter. Additionally, NJL models rely on phenomenological cutoff parameters, which introduce uncertainties into their predictions. Future constraints from high-energy heavy-ion collisions and neutron star observations will be essential for refining NJL-based quark matter models.
	
	\subsubsection{Functional Renormalization Group (FRG) Approach}
	
	The Functional Renormalization Group (FRG) approach is a more advanced method for studying quark matter, incorporating non-perturbative QCD effects that go beyond mean-field approximations. The FRG framework systematically accounts for fluctuations in quark interactions, leading to more accurate predictions of quark matter properties under extreme conditions. The basic assumptions of this model are as follows. First, non-perturbative QCD corrections are necessary to describe quark matter at high densities. Then, renormalization group flow equations determine how quark interactions evolve with density. Thirdly, color superconductivity phases are possible, altering the cooling and transport properties of neutron stars. In addition, effective density-dependent interactions replace fixed interaction strengths seen in NJL models.
	
	One of the advantages of the FRG approach is its ability to track the evolution of QCD coupling constants as a function of density, allowing for a more detailed description of phase transitions. Unlike NJL models, which assume fixed interaction strengths, FRG-based models provide a density-dependent interaction profile, leading to a more dynamically evolving equation of state.
	
	FRG-based quark matter models have been particularly useful in predicting color superconducting phases, where quarks form Cooper pairs similar to electrons in superconductors. These phases could significantly alter neutron star cooling rates, transport properties, and even the stability of quark matter cores. Recent studies using FRG techniques have suggested that strong repulsive interactions at high densities may help quark matter support neutron stars above the \(2 M_{\odot}\) threshold, making it a promising approach for understanding massive neutron stars.
	
	However, challenges remain in applying FRG methods to neutron star modeling. The computational complexity of FRG calculations makes it difficult to produce fully self-consistent neutron star equations of state, and further advancements in numerical QCD techniques are required to refine these predictions.
	
	\subsubsection{Holographic QCD (V-QCD) Models}
	
	Holographic QCD models, particularly V-QCD (Veneziano-type QCD) models, provide an alternative perspective on quark matter by utilizing the gauge-gravity duality from string theory. These models treat strongly coupled QCD matter as a gravitational system in a higher-dimensional spacetime, offering insights into quark interactions that are difficult to capture using traditional field theory approaches. The basic assumptions of this model are as follows. First, gauge-gravity duality provides an effective description of strongly coupled quark matter.Then, quark deconfinement occurs dynamically, leading to a possible stiffening of the EoS. Thirdly, strange quark matter could be absolutely stable, allowing for strange quark stars. Additionally, Holographic energy scales are used to determine effective QCD interactions at high densities.
	
	One of the key strengths of holographic QCD models is their ability to describe strongly interacting quark matter in a way that avoids the limitations of perturbative QCD and mean-field models. V-QCD models naturally predict a stiffer equation of state at high densities, making them compatible with observed neutron star mass constraints. Additionally, these models allow for the exploration of quark deconfinement mechanisms in a more controlled theoretical framework.
	
	A major prediction of V-QCD models is the existence of stable strange quark stars, where deconfined strange quark matter is absolutely stable and constitutes the entire star. This contrasts with hybrid models, where quark matter exists only in the inner core of neutron stars. Identifying whether strange quark stars exist would have profound implications for astrophysics, as their mass-radius relationships differ significantly from those of traditional neutron stars.
	
	Despite their promise, holographic QCD models face several theoretical and computational challenges. The translation of gauge-gravity results into physically meaningful QCD predictions remains an open problem, and fine-tuning these models to match empirical nuclear physics data is still an area of active research. Nonetheless, V-QCD models provide a valuable tool for exploring the strong coupling regime of QCD, which is crucial for understanding extreme astrophysical environments.
	
	\subsubsection{Key Conclusions}
	
	Quark matter models\cite{ref7} describe neutron stars as being composed entirely of deconfined quarks. These models generally predict softer equations of state, requiring additional mechanisms to support high-mass neutron stars.
	
	The Nambu–Jona-Lasinio (NJL) model describes quark matter interactions through chiral symmetry breaking and restoration mechanisms. The pressure in NJL models is given by:
	
	\begin{equation}
		P_{\text{NJL}} = -\frac{1}{V} \left( \Omega_{\text{quark}} + \Omega_{\text{interaction}} \right)
	\end{equation}
	
	where \(\Omega_{\text{quark}}\) accounts for the quark kinetic energy and \(\Omega_{\text{interaction}}\) includes four-fermion interaction terms. NJL models often predict neutron stars with maximum masses below \(2 M_{\odot}\), making them inconsistent with observed high-mass pulsars unless additional repulsive interactions are included.
	
	In the Functional Renormalization Group (FRG) approach, QCD interactions evolve dynamically with density. This approach introduces a density-dependent quark-quark interaction term governed by the renormalization group flow equation:
	
	\begin{equation}
		\frac{d \Gamma_k}{dk} = \frac{1}{2} \text{Tr} \left[ \frac{\partial_k R_k}{\Gamma_k^{(2)} + R_k} \right]
	\end{equation}
	
	where \(\Gamma_k\) is the effective action at a given renormalization scale. FRG models suggest that strong repulsive interactions at high densities can make quark matter compatible with high-mass neutron stars.
	
	Finally, holographic QCD (V-QCD) models use gauge-gravity duality to describe strongly interacting quark matter. The energy density in these models is given by:
	
	\begin{equation}
		\epsilon = \frac{3}{4} \frac{N_c^2}{\lambda} T^4
	\end{equation}
	
	where \(N_c\) is the number of colors, \(\lambda\) is the QCD coupling parameter, and \(T\) is the temperature. V-QCD models predict very stiff equations of state, often supporting neutron stars with masses above \(2.2M_{\odot}\). Some V-QCD models suggest the existence of stable strange quark stars, where deconfined strange quark matter is absolutely stable, leading to distinct mass-radius relationships compared to traditional neutron stars.
	
	Quark matter models remain an active area of research, with future astrophysical and experimental constraints expected to further refine their viability.
	
	\subsection{Hybrid Models}
	
	Hybrid models incorporate a transition from hadronic matter to deconfined quark matter at sufficiently high densities, providing a more complete description of neutron star interiors. These models bridge the gap between purely hadronic descriptions and full quark matter equations of state by allowing for the coexistence of nucleonic and quark phases within a neutron star core. The transition between hadronic and quark matter can occur in different ways, with some models predicting a first-order phase transition and others favoring a smooth crossover transition. The nature of this transition has profound implications for neutron star properties, including mass, radius, and tidal deformability.
	
	\subsubsection{Quark-Hadron Crossover (QHC) Models}
	
	One class of hybrid models is the Quark-Hadron Crossover (QHC) Model, which describes a gradual transition from hadronic matter to quark matter rather than a sharp phase boundary. These models are motivated by lattice QCD simulations, which suggest that at high densities, quark degrees of freedom emerge continuously rather than through a sudden phase transition. The QHC models provide an intermediate stiffness in the equation of state, balancing the pressure contributions from both hadronic and quark matter components. The basic assumptions of this model are as follows. First, the transition from hadronic matter to quark matter occurs gradually, avoiding a sharp phase boundary. Then, Density-dependent interactions govern the emergence of quark degrees of freedom at high densities. Thirdly, Chiral symmetry restoration happens smoothly as quark matter forms. Additionally, Beta equilibrium and charge neutrality are maintained throughout the transition.
	
	An example of a well-studied QHC model is QHC21\cite{ref8}, which smoothly connects hadronic matter at lower densities with quark matter at extreme densities. This model is tuned to astrophysical constraints from NICER observations of neutron star radii and gravitational wave events such as GW170817. One of the key advantages of QHC models is their ability to support high-mass neutron stars around 2.1 $M_{\odot}$ while maintaining consistency with observed tidal deformability constraints from neutron star mergers. The phase transitions under different parameters for the hybrid models are shown in Figure 4 adapted from\cite{ref7}.
	\begin{figure}[H]
		\centering
		\includegraphics[width=0.75\textwidth]{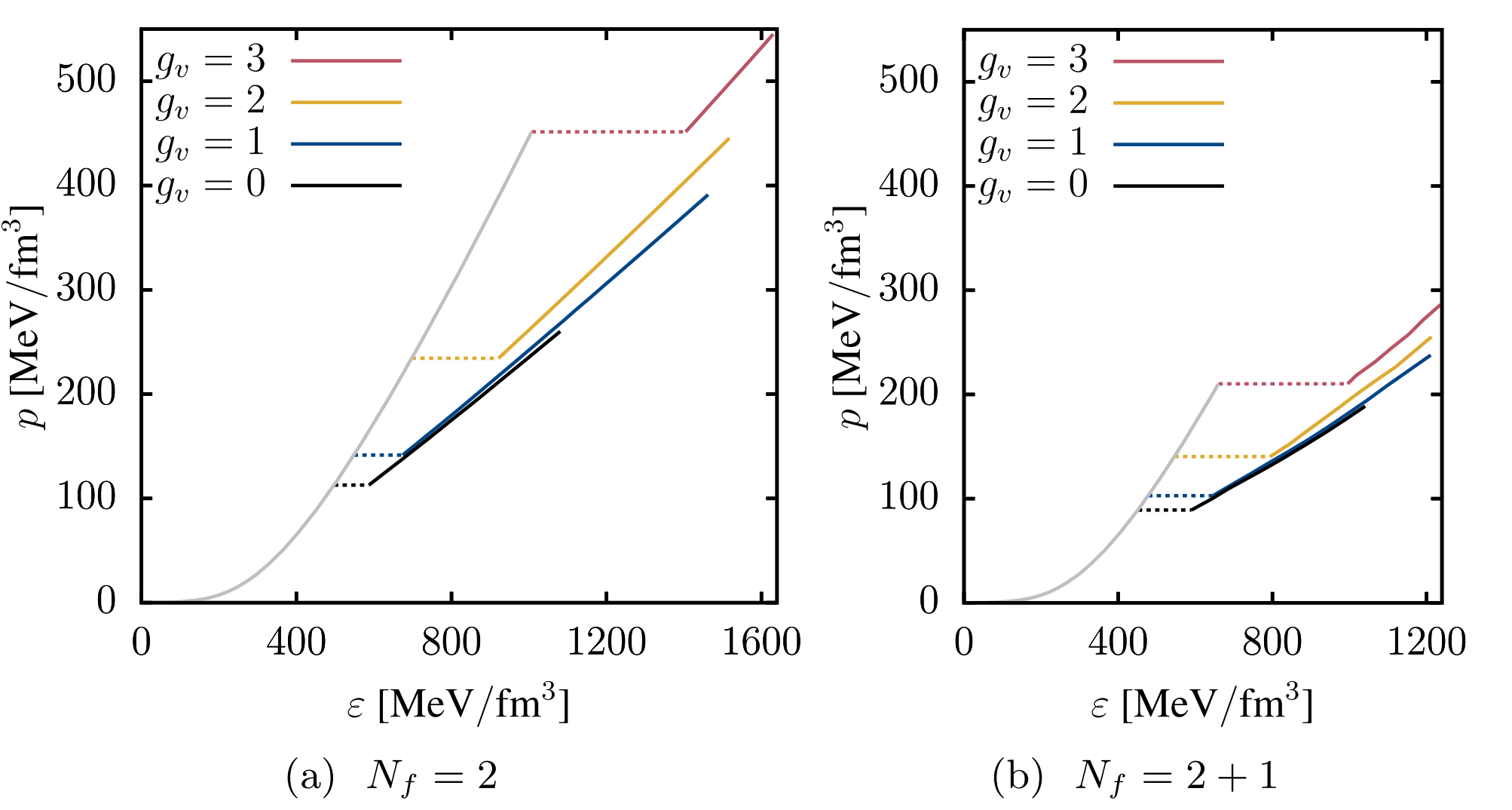}
		\caption{Equation of state for hybrid matter. The nucleonic phase is described by the HS(DD2) EoS (gray color), and the quark matter phase is represented by the FRG quark-meson EoS with vector mesons. Both phases separately satisfy weak equilibrium and charge neutrality and are connected via a Maxwell construction. For \( N_f = 2 + 1 \), energy densities larger than \( \varepsilon \sim 1200\ \mathrm{MeV/fm^3} \), corresponding to \( \mu > 500\ \mathrm{MeV} \), are dropped.
		}
		\label{fig4}
	\end{figure}
	
	Unlike traditional hadronic models, which become excessively stiff at high densities, QHC models introduce density-dependent quark interactions that moderate the pressure increase, preventing unrealistic mass predictions. However, one of the challenges with QHC models is the lack of direct experimental verification for quark-hadron crossover behavior at high densities. Future observations of neutron stars with extreme masses or precise gravitational wave detections from post-merger oscillations could provide crucial constraints on the validity of these models.
	
	\subsubsection{First-Order Phase Transition Models}
	
	Another widely studied category of hybrid models involves first-order phase transitions, where a sharp transition from hadronic matter to a deconfined quark phase occurs at a critical density\cite{ref9}. This transition is often modeled using either the Maxwell construction, which assumes a sharp density discontinuity, or the Gibbs construction, which allows for a mixed-phase region where hadronic and quark matter coexist. The basic assumptions of this model are as follows. First, a sudden phase transition occurs at a critical density, resulting in a distinct hadronic-quark matter boundary. Then, a sharp pressure discontinuity is present in the Maxwell construction, while the Gibbs construction allows a mixed phase. Thridly, Quark matter is more stable at high densities than hadronic matter. In addition, the existence of quark cores inside massive neutron stars is possible.
	
	First-order phase transition models predict distinct structural changes in neutron stars, particularly in the form of a quark core surrounded by a hadronic envelope. If the phase transition occurs at moderate densities, around 2–4 times nuclear saturation density, it can lead to the formation of mass twins—neutron stars with the same mass but different internal compositions, depending on whether they contain a quark core. This phenomenon has been explored as a potential explanation for variations in neutron star radius measurements.
	
	Observational constraints from gravitational wave detections, particularly from binary neutron star mergers like GW170817, have played a significant role in shaping first-order transition models\cite{ref10}. The measured tidal deformability from such events suggests that excessively soft equations of state are unlikely, meaning that phase transitions must occur at sufficiently high densities to avoid conflict with astrophysical observations. Future gravitational wave events and potential detections of post-merger oscillations could provide further evidence for (or against) the existence of first-order phase transitions in neutron stars.
	
	One of the key challenges of first-order phase transition models is the uncertainty in the transition density and pressure\cite{ref11}. While some models predict that the transition occurs at relatively low densities, making quark cores a common feature in neutron stars, others suggest that it happens only at extremely high densities, limiting the presence of quark matter to the most massive neutron stars. Current and future astrophysical observations will be critical in refining these models and determining whether quark cores are a common feature in neutron star populations.
	
	\subsubsection{Key Conclusions}
	
	Hybrid models incorporate both hadronic and quark matter components, allowing for a transition between them. The key difference among hybrid models is the nature of the transition—either a smooth crossover or a first-order phase transition.
	
	In quark-hadron crossover (QHC) models, the transition between hadronic and quark matter is gradual rather than abrupt. These models are motivated by lattice QCD simulations, which suggest that chiral symmetry restoration in QCD can occur as a smooth crossover rather than a sharp phase transition. The crossover is typically parameterized using a weighted pressure function\cite{ref8}:
	
	\begin{equation}
		P_{\text{QHC}}(\epsilon) = w P_{\text{hadronic}} + (1-w) P_{\text{quark}}
	\end{equation}
	
	where \(w(\rho)\) smoothly interpolates between the two phases. QHC models successfully explain high-mass neutron stars (\(\sim2.1M_{\odot}\)) while maintaining consistency with NICER radius constraints and tidal deformability constraints from GW170817.
	
	In contrast, first-order phase transition models\cite{ref9} predict a sharp transition from hadronic matter to quark matter at a critical pressure \(P_c\). This transition is modeled using the Maxwell construction, which assumes that pressure remains constant across the phase boundary:
	
	\begin{equation}
		P_{\text{hadronic}} = P_{\text{quark}}, \quad \mu_{\text{hadronic}} = \mu_{\text{quark}}
	\end{equation}
	
	Alternatively, the Gibbs construction allows for a mixed-phase region, where hadronic and quark matter coexist, and the pressure is given by a weighted sum of the two phases:
	
	\begin{equation}
		P_{\text{mix}} = \chi P_{\text{hadronic}} + (1-\chi) P_{\text{quark}}
	\end{equation}
	
	where \(\chi\) is the volume fraction of quark matter. First-order phase transitions can lead to mass twins—neutron stars with the same mass but different internal compositions depending on whether they contain a quark core. The presence of a phase transition can also affect gravitational wave signals from neutron star mergers, producing post-merger oscillations that may be detectable by next-generation gravitational wave observatories.
	
	\subsection{Model Based on Chiral EFT}
	\subsubsection{MBPT Model}
	
	In the theoretical modeling of the properties of dense matter inside neutron stars, Chiral Effective Field Theory ($\chi$EFT) provides a systematic and controlled approach to constructing nuclear forces\cite{ref12,ref13}. This theoretical framework employs low-energy effective degrees of freedom—primarily nucleons and pions—and expands interactions order by order in momentum scale and cutoff. Through this expansion, two-body (NN), three-body (3N), and even four-body (4N) interactions can be incorporated step by step, allowing for the construction of high-precision equations of state (EOS) for neutron matter. Within this framework, recent studies have not only developed complete EOS models up to next-to-next-to-next-to-leading order (N3LO) but also introduced Bayesian statistical methods to quantify truncation uncertainties, thereby enhancing the predictive power and reliability of the models.
	
	The EOS models constructed under this framework are based on many-body perturbation theory (MBPT), which involves systematic calculations of the neutron matter energy, pressure, and sound speed through Hartree-Fock, second-order, and third-order perturbative expansions. The model assumes zero-temperature pure neutron matter (PNM) as the physical state inside neutron stars and incorporates symmetric nuclear matter (SNM) to model the symmetry energy component. The nuclear interactions employed are derived from $\chi$EFT-based NN potentials with various cutoff parameters (such as EGM and EM potentials), while the low-energy constants (LECs) of the three-body forces, such as $c_1$ and $c_3$, are chosen based on theoretical analyses and empirical saturation properties. In advanced EOS models, three-body and four-body forces are fully included at N3LO order, which is particularly crucial for describing the equation of state in the high-density regime of neutron stars.
	
	To overcome the limitations of conventional $\chi$EFT models in handling truncation uncertainties, recent studies further incorporate Gaussian Process (GP) regression as a statistical learning tool. This method is trained on existing $\chi$EFT calculations at different orders and automatically learns the density dependence and covariance structure of the uncertainties across physical quantities. Through multitask learning, the model is capable of capturing the correlated truncation uncertainties between PNM and SNM, thus improving the reliability of predictions for derived quantities such as the symmetry energy and its density slope. The implementation of GP also enables the estimation of probability distributions for physical quantities such as pressure and sound speed without introducing additional model parameters.
	
	\subsubsection{Key Conclusions}
	
	This model develops a novel Bayesian uncertainty quantification (UQ) framework to constrain the neutron-star equation of state (EOS) at densities up to twice the nuclear saturation density. The approach combines many-body perturbation theory (MBPT) calculations with chiral effective field theory interactions, while accounting for correlated truncation errors both across density and between physical observables. Through the use of Gaussian Process (GP) regression with physics-informed hyperparameters, we achieve statistically robust and smoothly propagating uncertainties for all derived EOS quantities\cite{ref12,ref13}.
	
	The predictions obtained for key nuclear matter parameters at saturation density $n_0 \approx 0.16~\mathrm{fm}^{-3}$ are in excellent agreement with experimental constraints. In particular, the symmetry energy and its slope are estimated as:
	\begin{align}
		S_v &= S_2(n_0) = 31.7 \pm 1.1~\mathrm{MeV}, \\
		L &= L(n_0) = 59.8 \pm 4.1~\mathrm{MeV}.
	\end{align}
	Moreover, the squared speed of sound at twice saturation density remains below the conformal limit,
	\begin{equation}
		c_s^2(2n_0) \approx 0.10 \pm 0.07~(\text{at N3LO}),
	\end{equation}
	though higher-density behavior may exceed this value, as suggested by astrophysical observations of neutron stars with masses exceeding $2~M_\odot$.
	
	A key advancement of this framework lies in its ability to incorporate strong correlations between observables. Notably, the energy per particle in pure neutron matter, $E/N(n)$, and symmetric nuclear matter, $E/A(n)$, are found to be highly correlated, with a Pearson correlation coefficient of
	\begin{equation}
		\rho_{E/N,~E/A} \approx 0.94.
	\end{equation}
	
	The posterior distribution of $(S_v, L)$ obtained from the GP regression is well-approximated by a bivariate Gaussian, with mean and covariance matrix:
	\begin{align}
		\boldsymbol{\mu} &= \begin{bmatrix} 31.7 \\ 59.8 \end{bmatrix}~\mathrm{MeV}, \\
		\boldsymbol{\Sigma} &= \begin{bmatrix}
			1.112 & 3.27 \\
			3.27 & 4.122
		\end{bmatrix}~\mathrm{MeV}^2.
	\end{align}
	The resulting credible regions in the $(S_v, L)$ plane lie within the overlap area of constraints from experimental data such as dipole polarizability, neutron-skin thickness, and heavy-ion collisions, indicating excellent consistency with empirical observations. Figure 8 from \cite{ref12} presents some of the predicted results from this model.
	
	Unlike traditional EFT error prescriptions, this framework allows for smooth propagation of uncertainties into derivatives and complex observables. Furthermore, the method is highly extensible. It can be directly adapted for finite-temperature equations of state, arbitrary isospin asymmetry, and even full Bayesian inference using Markov Chain Monte Carlo sampling over low-energy constants and GP hyperparameters. Its compatibility with multimessenger observational data, such as from NICER and LIGO/Virgo, makes it a powerful tool for future EOS reconstruction.
	
	\begin{figure}[H]
		\centering
		\subfigure[Energy]{
			\includegraphics[width=0.42\textwidth]{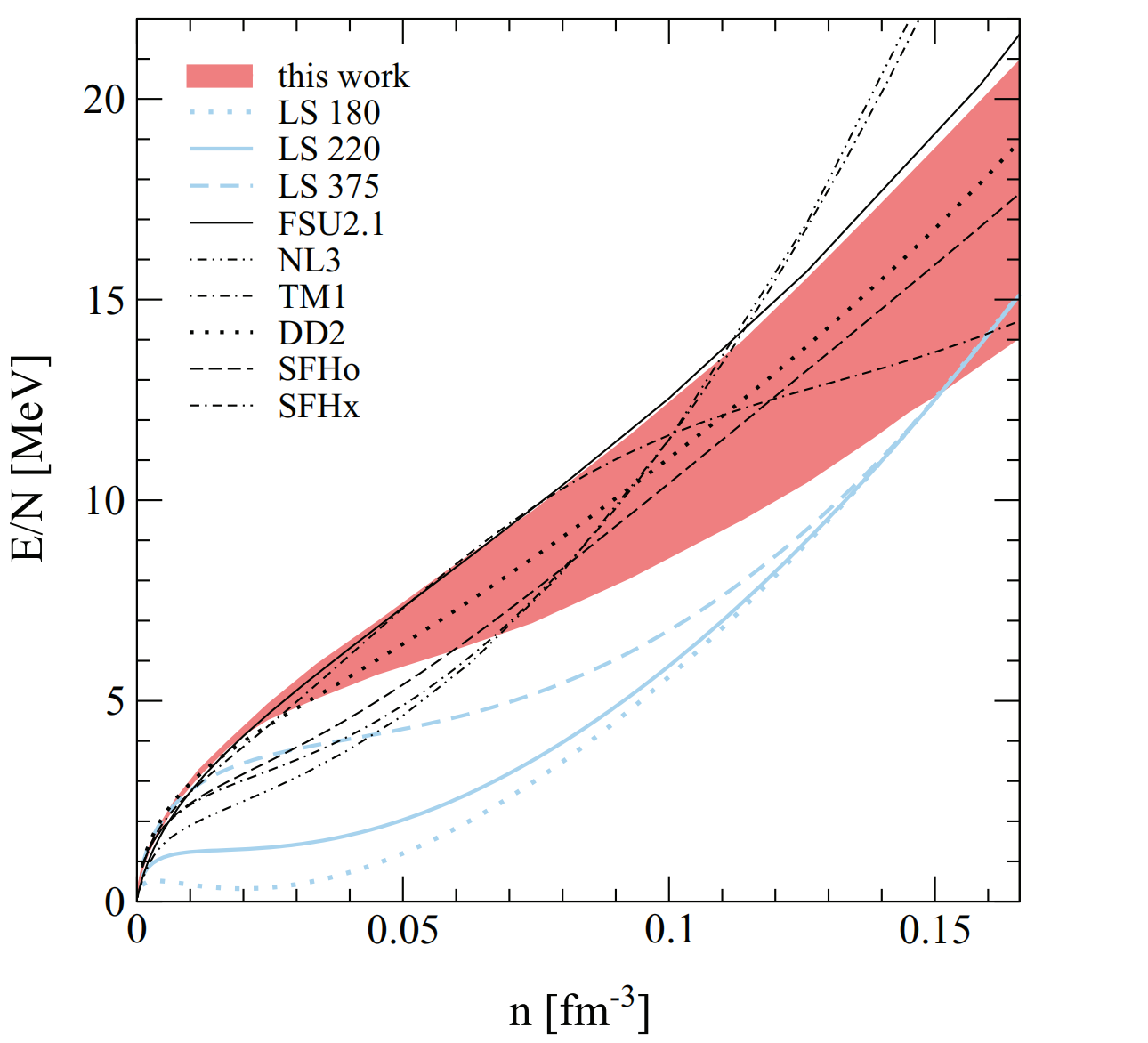}
		}
		\hspace{0.0\textwidth}
		\subfigure[Radius–mass]{
			\includegraphics[width=0.4\textwidth]{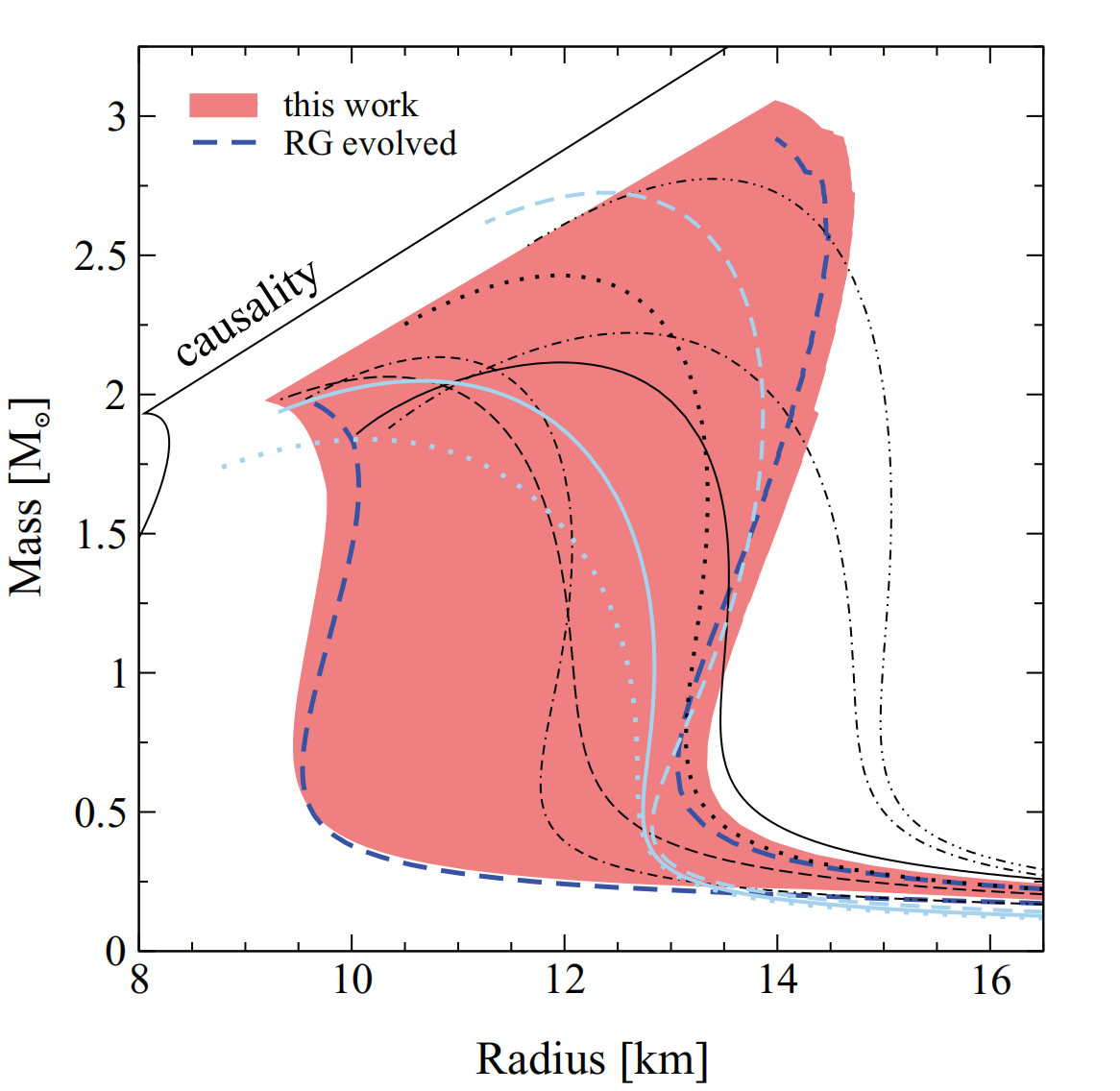}
		}
		\caption{(a) Comparison of the neutron-matter energy (red band) with equations of state for core-collapse supernova simulations (b) Radius–mass relationship of the Chiral EFT model at N$^3$LO}
		\label{fig8}
	\end{figure}

	\section{Constraints on the Equation of State}
	\label{3}
	The equation of state (EoS) of neutron stars is subject to stringent constraints from multiple independent sources, including astrophysical observations, nuclear physics experiments, and theoretical quantum chromodynamics (QCD) predictions\cite{ref14}. These constraints refine EoS models by limiting the range of acceptable stiffness, phase transitions, and compositions. In this section, we systematically discuss the major constraints imposed on the neutron star EoS and present the essential equations and relationships that reflect neutron star structure.
	
	\subsection{Mass and Radius Constraints}
	
	Observations of neutron star masses and radii provide some of the strongest constraints on the EoS. The Tolman-Oppenheimer-Volkoff (TOV) equation\cite{ref15}, derived from general relativity, describes the balance between gravitational collapse and internal pressure support in a neutron star:
	
	\begin{equation}
		\frac{dP}{dr} = -\frac{G(\epsilon + P)(m + 4\pi r^3 P)}{r^2 (1 - 2Gm/r)}
	\end{equation}
	
	where \( P(r) \) is the pressure at radius \( r \), \( \epsilon(r) \) is the energy density, \( m(r) \) is the mass enclosed within radius \( r \), and \( G \) is the gravitational constant.
	
	\begin{figure}[H]
		\centering
		\includegraphics[width=0.55\textwidth]{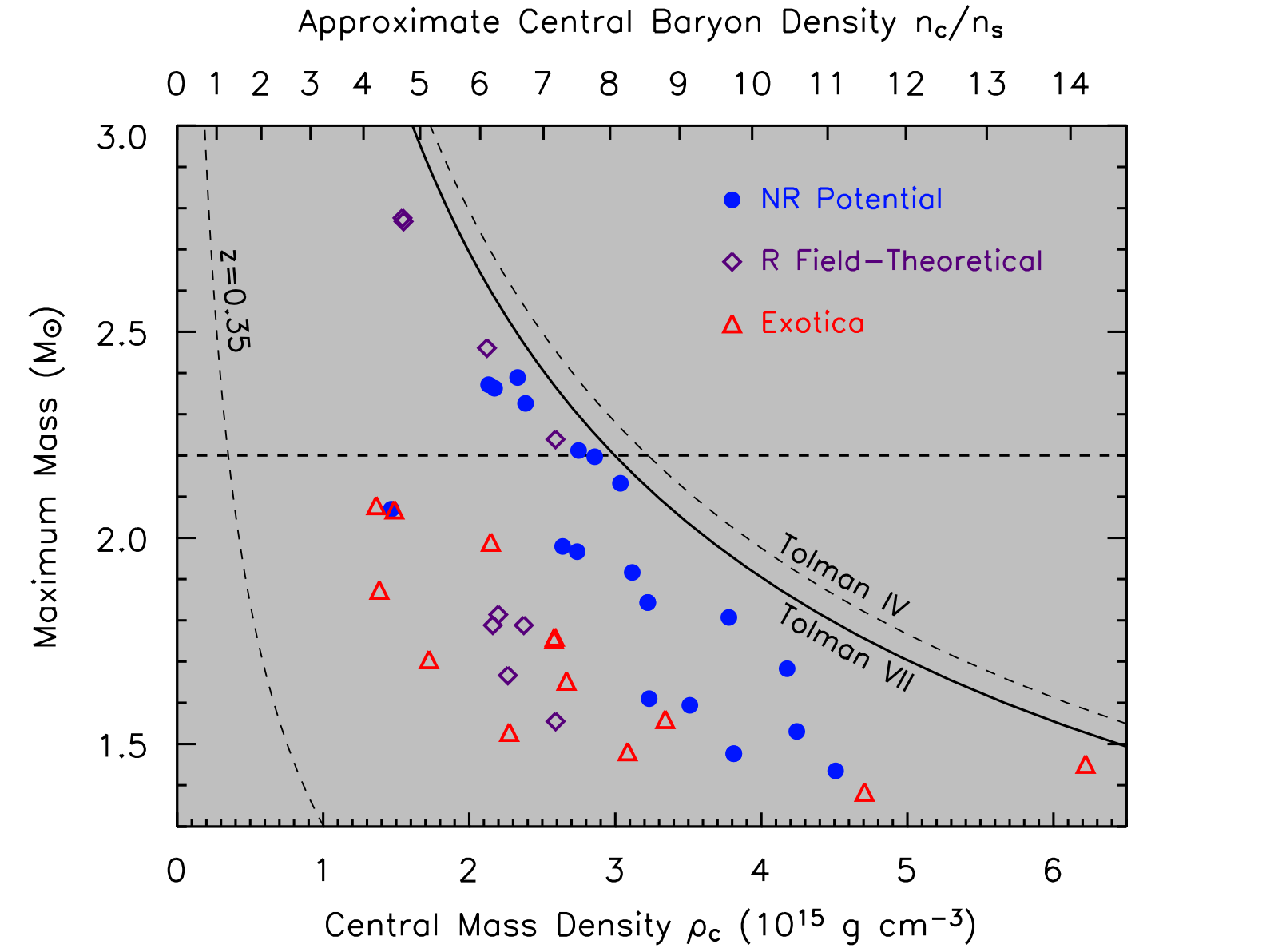}
		\caption{The maximum mass–central density relation. Figure taken from\cite{ref22}}
		\label{fig5}
	\end{figure}
	
	The maximum mass–central density relation predicted by causality coupled with the Tolman VII and Tolman IV analytic GR solutions are compared with structure integration results for a variety of EOSs in figure 6. NR refers to non-relativistic potential EOSs, R refers to relativistic field-theoretical EOSs, and Exotica refers to EOSs with considerable softening at high density due to kaon condensation or strange quark matter deconfinement. A possible redshift measurement of \( z \) = 0.35 is also shown.
	
	Observations of high-mass pulsars such as PSR J1614-2230 (\( M = 1.97 \pm 0.04 M_{\odot} \)) and PSR J0740+6620 (\( M = 2.08 \pm 0.07 M_{\odot} \))\cite{ref16} confirm that neutron stars must be able to support at least \(2.0 M_{\odot}\) against gravitational collapse. Additionally, X-ray timing observations from NICER have constrained the radius of a \(1.4 M_{\odot}\) neutron star to be:
	
	\begin{equation}
		12.2 \text{ km} \leq R_{1.4} \leq 13.7 \text{ km}
	\end{equation}
	
	which rules out equations of state that are too soft (leading to small radii) or too stiff (predicting excessively large neutron stars). A clearer relationship is presented in Figure 7 from Ref\cite{ref21}. Additional constraints from PSR J0740+6620 and PSR J1614-2230 were taken into account.
	
	\begin{figure}[H]
		\centering
		\includegraphics[width=0.8\textwidth]{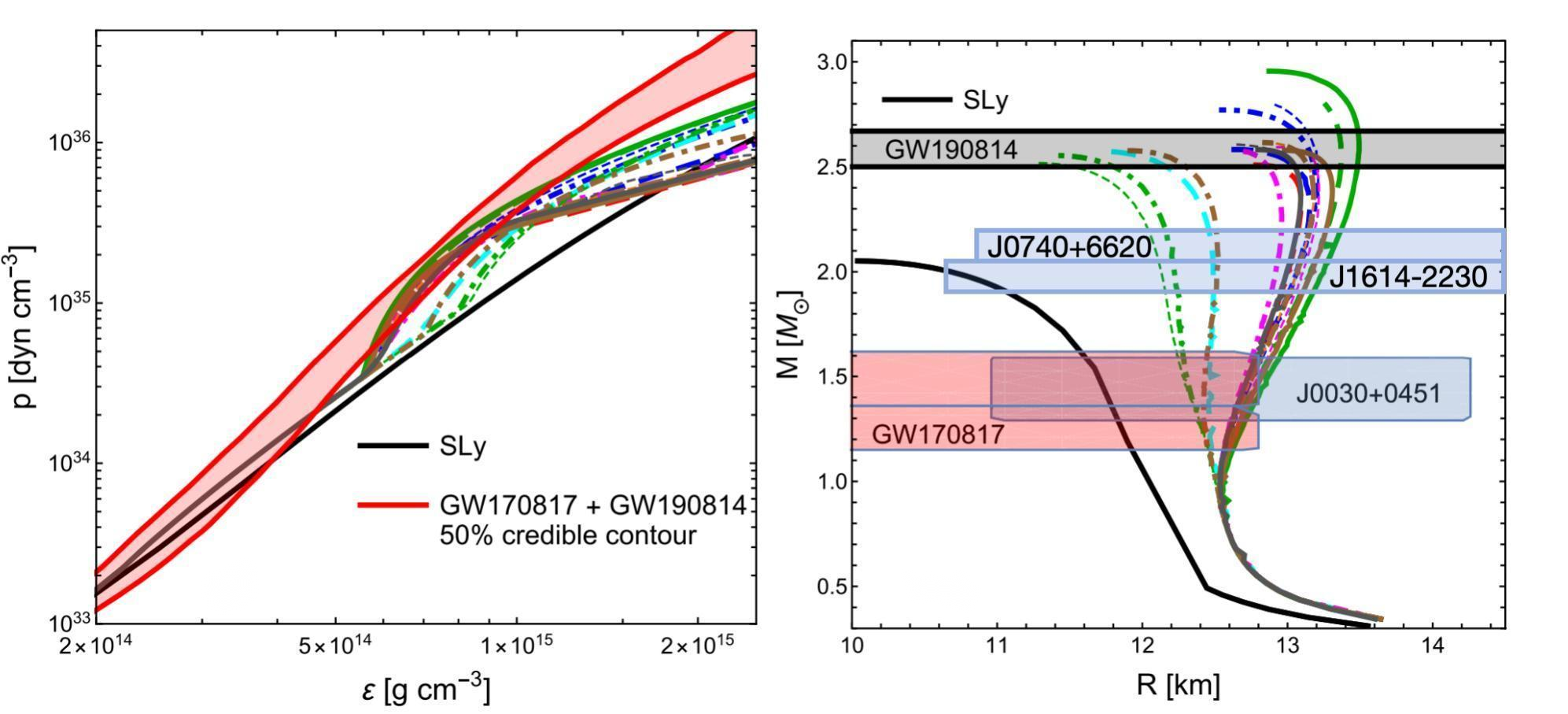}
		\caption{The left panel shows the equations of state (EOSs) resulting from the parameterized speed of sound yield the relation between pressure \( p \) and energy density \( \varepsilon \). The red shaded region corresponds to the 90\% credible interval derived from the joint analysis of GW170817 and GW190814, while the red curve indicates the 50\% credible contour. The black solid curve represents the SLy EOS used as a reference. 
			The right panal shows the mass–radius (M–\( R \)) relations resulting from the new EOSs. It can be seen that all neutron star sequences reach a maximum mass of at least \( M_{\mathrm{max}} \geq 2.5~M_\odot \). The gray band denotes the inferred mass of the secondary compact object in GW190814, and the blue rectangle indicates the mass and radius constraints obtained from NICER observations of the isolated pulsar PSR J0030+0451 and PSR J0740+6620, with the PSR J1614-2230 from Arecibo observatory.
		}
		\label{fig6}
	\end{figure}
	In Figure 7, different colored curves represent distinct EOS families with specific features in the speed of sound. Green curves correspond to models where \( c_s^2 \) sharply rises to the causal limit at varying densities. Cyan curves use a hyperbolic tangent to transition smoothly to a constant \( c_s^2 \). Blue curves vary the asymptotic value of \( c_s^2 \) at high densities. Orange curves shift the location of the speed of sound peak. Brown curves vary both the peak position and width to ensure \( M_{\text{max}} \geq 2.5\,M_\odot \). Dark gray curves share the same initial peak but differ in the functional form beyond it. Magenta curves introduce oscillations after the initial rise. Red curves feature a double-peak structure. These families explore different physical scenarios consistent with current astrophysical constraints.
	
	\subsection{Tidal Deformability Constraints}
	
	Gravitational wave detections of binary neutron star mergers provide additional constraints on the EoS through tidal deformability measurements\cite{ref16}. The tidal deformability parameter \( \Lambda \) quantifies how much a neutron star deforms under the gravitational field of its companion. It is given by:
	
	\begin{equation}
		\Lambda = \frac{2}{3} k_2 \left(\frac{R}{M}\right)^5
	\end{equation}
	
	where \( k_2 \) is the Love number, \( R \) is the neutron star radius, and \( M \) is its mass\cite{ref10}.
	
	From GW170817, the LIGO/Virgo analysis constrained the tidal deformability of a \(1.4 M_{\odot}\) neutron star as:
	
	\begin{equation}
		70 \leq \Lambda_{1.4} \leq 580
	\end{equation}
	
	This result ruled out overly soft equations of state that predict high tidal deformabilities. Future gravitational wave detections (e.g., Einstein Telescope, Cosmic Explorer\cite{ref14}) will further refine these constraints. As shown in Figure 8\cite{ref10}, not all models meet this constraint.
	
	\begin{figure}[H]
		\centering
		\includegraphics[width=0.75\textwidth]{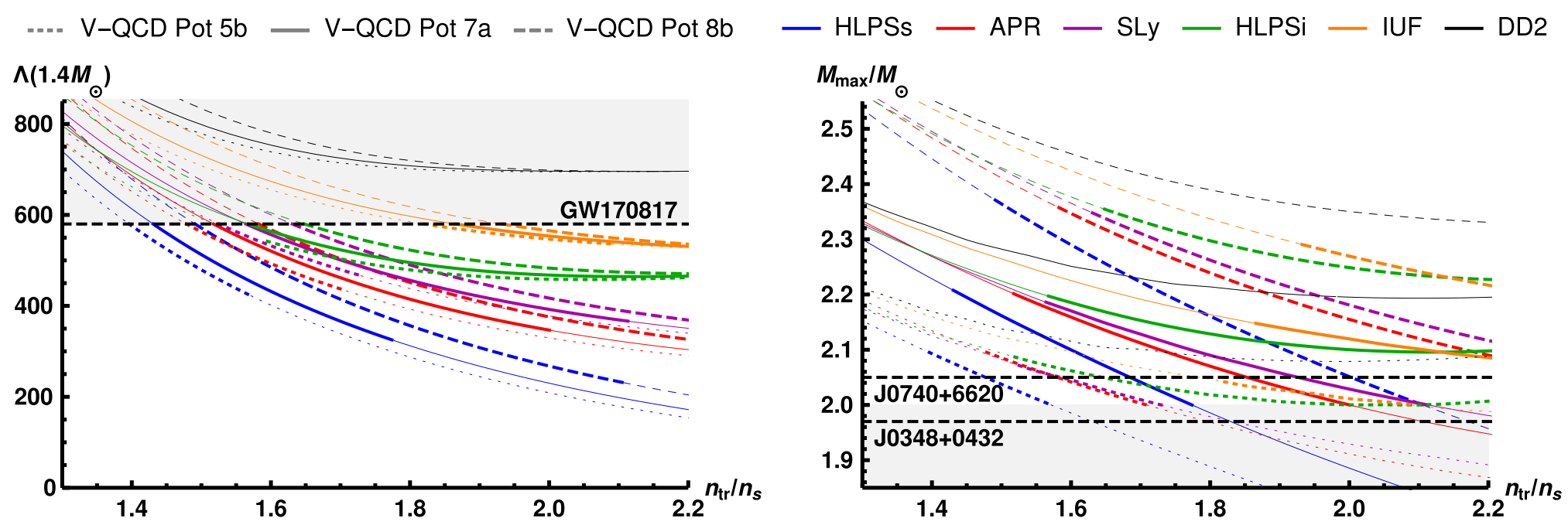}
		\caption{Constraints on the range of tidal deformability from observational results}
		\label{fig7}
	\end{figure}
	Figure 8 presents observational constraints on the tidal deformability and maximum mass of neutron stars as functions of the transition density normalized to nuclear saturation density, \( n_T/n_s \). In the left panel, the horizontal axis represents \( n_T/n_s \), while the vertical axis shows the dimensionless tidal deformability \( \Lambda(1.4\,M_\odot) \) for a canonical \( 1.4\,M_\odot \) neutron star. The shaded region and the dashed line indicate the upper limit on \( \Lambda \) from the GW170817 event. In the right panel, the vertical axis represents the maximum mass \( M_{\text{max}}/M_\odot \) predicted by each EOS, with shaded areas corresponding to observed lower bounds from heavy pulsars such as J0740+6620 and J0348+0432. Together, these plots constrain the viable range of EOSs by requiring consistency with both tidal deformability and mass measurements.

	\subsection{Constraints from Nuclear Physics Experiments}
	
	Laboratory nuclear physics experiments provide constraints on the low-density behavior of the neutron star EoS. These constraints come from neutron skin thickness measurements, heavy-ion collisions, and nuclear symmetry energy studies\cite{ref17}.
	
	The properties of symmetric nuclear matter at saturation density (\(\rho_0 \approx 0.16\) fm\(^{-3}\)) provide an anchor for neutron star EoS models. Key empirical nuclear saturation properties include:
	
	\begin{equation}
		E_{\text{sat}} \approx -16 \text{ MeV}, \quad K_{\text{sat}} \approx 230 \text{ MeV}, \quad S_0 \approx 30-35 \text{ MeV}
	\end{equation}
	
	Neutron skin thickness measurements, such as those from the PREX-II experiment\cite{ref18}, provide insights into the nuclear symmetry energy, indirectly constraining neutron star radii. Heavy-ion collision experiments at GSI, RIKEN, FAIR, and NICA\cite{ref21} probe nuclear matter at densities of \(2-4 \rho_0\), similar to neutron star interiors. These experiments suggest that the pressure at twice nuclear saturation density should be in the range:
	
	\begin{equation}
		P(2\rho_0) \approx 50-80 \text{ MeV/fm}^3
	\end{equation}
	
	Any EoS that deviates significantly from this range is inconsistent with terrestrial nuclear data.
	
	\subsection{Theoretical Constraints from QCD and High-Density Matter}
	
	At ultra-high densities (\(\rho > 5 \rho_0\)), neutron stars probe the non-perturbative regime of QCD. Several theoretical approaches provide constraints on the high-density EoS\cite{ref20}.
	
	At asymptotically high densities, perturbative QCD predicts that the pressure \( P \) follows:
	
	\begin{equation}
		P \sim \mu^4
	\end{equation}
	
	where \( \mu \) is the quark chemical potential. This requires that quark matter must be sufficiently stiff to support neutron stars with \( M > 2 M_{\odot} \).
	
	Strange quark matter (SQM) models suggest that deconfined quark matter could be absolutely stable, leading to the existence of strange quark stars. The stability condition is given by:
	
	\begin{equation}
		E_{\text{SQM}} / A \leq 930 \text{ MeV}
	\end{equation}
	
	If this condition holds, neutron stars could actually be strange quark stars rather than hadronic stars.
	
	\subsection{Mass-Radius and EoS Comparisons}
	
	Different equations of state (EoS) provide distinct predictions for neutron star properties, including maximum mass, radius, tidal deformability, and phase transitions. These predictions can be systematically compared to evaluate their viability against observational and experimental constraints. By combining mass, radius, tidal deformability, and theoretical constraints, different EoS models can be compared systematically\cite{ref3,ref4,ref5,ref6,ref7,ref8,ref9,ref10,ref11,ref12,ref13,ref23,ref25}.
	
	\begin{table}[h]
		\centering
		\caption{Predictions of Maximum Mass, Radius, and Tidal Deformability for Different EoS Models}
		\begin{tabular}{lccc}
			\hline
			\textbf{Model Type} & \textbf{Maximum Mass} (\(M_{\odot}\)) & \textbf{Radius at \(1.4 M_{\odot}\)} (km) & \textbf{Tidal Deformability} (\(\Lambda_{1.4}\)) \\
			\hline
			RMF (GM1, DD2) & \(2.0 - 2.2\) & \(12.0 - 13.0\) & \(200 - 500\) \\
			Hyperonic RMF & \(1.6 - 1.9\) & \(11.5 - 12.5\) & \(>500\) \\
			Kaon Condensation & \(1.8 - 2.0\) & \(11.5 - 12.3\) & \(>500\) \\
			QHC21 & \(2.1\) & \(12.4\) & \(250 - 450\) \\
			Hybrid Model & \(2.0 - 2.1\) & \(12.0 - 12.6\) & \(150 - 400\) \\
			NJL & \(<2.0\) & \(11.0 - 11.8\) & \(>500\) \\
			FRG & \(>2.1\) & \(12.5 - 13.0\) & \(150 - 350\) \\
			V-QCD & \(>2.2\) & \(12.8 - 13.2\) & \(100 - 300\) \\
			\hline
		\end{tabular}
		\label{tab:eos_predictions}
	\end{table}
	
	RMF models predict stiff equations of state that can support high-mass neutron stars, while hyperonic and kaon-condensation models often fail to explain observed \(2 M_{\odot}\) pulsars. Hybrid models and stiff quark matter models, such as FRG and V-QCD, generally provide mass and radius predictions that align with observations\cite{ref22}.
	
	Hybrid EoS models incorporating a phase transition can produce mass twins, neutron stars with the same mass but different internal compositions.
	
	\begin{table}[h]
		\centering
		\caption{Mass Twin Predictions in Different EoS Models}
		\begin{tabular}{lcc}
			\hline
			\textbf{Model Type} & \textbf{Mass Twin Effect}  & \textbf{Phase Transition} \\
			\hline
			RMF (GM1, DD2) & None & No \\
			QHC21 & Weak & Crossover \\
			FRG Approach & None & Smooth transition\\
			NJL & Moderate & First-order \\
			V-QCD & Strong & First-order\\
			\hline
		\end{tabular}
		\label{tab:mass_twins_comparison}
	\end{table}
	
	First-order hybrid models and V-QCD models predict strong mass twin effects, which could be confirmed through future precise neutron star mass-radius measurements.
	
	With upcoming astrophysical and experimental advancements, the constraints on the neutron star EoS will become even more stringent. Future gravitational wave detectors such as the Einstein Telescope (ET) and Cosmic Explorer (CE) will provide significantly improved constraints on neutron star tidal deformability\cite{ref10,ref14}, and the Laser Interferometer Space Antenna (LISA) is capable of detecting gravitational waves during the inspiral phase\cite{ref26}, helping distinguish between competing EoS models. Continued observations from NICER, combined with upcoming X-ray polarimetry missions like IXPE, will refine neutron star mass-radius measurements. Compared to NASA's previous flagship instrument RXTE (Rossi X-ray Timing Explorer), NICER achieves a sensitivity improvement of approximately 2 to 3 times in the key energy range (0.2–12 keV), owing to its enhanced effective area and time resolution\cite{ref27}. Meanwhile, IXPE's sensitivity to the X-ray polarization of celestial objects is improved by a factor of 10 to 100 compared to the earlier OSO-8 satellite, depending on the energy band and the target source\cite{ref28}. Further constraining the stiffness of the EoS. Laboratory studies at facilities such as FAIR in Germany, NICA in Russia, and J-PARC in Japan\cite{ref14} will explore high-density nuclear matter, providing critical input for nuclear models of the neutron star EoS.
	
	The combination of gravitational wave signals, electromagnetic counterparts, and neutrino detections from neutron star mergers will provide unprecedented constraints on neutron star composition, particularly regarding the presence of quark matter. As observational techniques and theoretical modeling improve, the neutron star EoS will become one of the most precisely constrained aspects of dense matter physics.
	
	\section{Future Prospects}
	\label{4}
	The study of neutron star equations of state (EoS) is rapidly advancing through progress in theoretical modeling, astronomical observations, and nuclear experiments. While substantial improvements have been made, key challenges remain. Future developments will focus on refining hadronic, hybrid, and quark matter models, particularly through the inclusion of three-body forces, density-dependent couplings, and more realistic QCD-based interactions such as FRG and holographic approaches. Machine-learning methods may also enhance EoS predictions.
	
	Advances in theoretical frameworks, including effective field theory and modified gravity extensions to the Tolman–Oppenheimer–Volkoff equation, will improve the description of dense matter. Enhanced lattice QCD and pQCD calculations will better constrain high-density quark matter, while neutrino transport and cooling models will further refine thermal evolution predictions.
	
	Observational constraints will benefit from next-generation facilities such as the Einstein Telescope, Cosmic Explorer, and SKA, enabling more precise mass-radius and tidal deformability measurements. Multimessenger astrophysics and potential detection of post-merger gravitational wave signatures or mass twins will offer critical insights into the existence of phase transitions and quark cores\cite{ref28}.
	
	Experimental progress from heavy-ion collisions not only from FAIR, NICA, J-PARC as mentioned above but also BEPCII in China\cite{ref30} and neutron skin measurements (e.g., PREX-II\cite{ref18}) will continue to constrain the symmetry energy and stiffness of nuclear matter. These data, combined with improved high-density nuclear interaction measurements, will help refine EoS models.
	
	Despite progress, fundamental questions remain, such as the existence of quark cores, the nature of mass twins, and the possibility of strange quark stars. Future studies must also consider magnetic fields, rotation, finite temperature, and potential dark matter interactions. A continued synergy between theory, experiment, and observation will be essential to fully understand the nature of ultra-dense matter in neutron stars.
	
	\section{Conclusion}
	\label{5}
	In this review, we have examined various equation of state (EoS) models for neutron stars, highlighting their theoretical basis, key predictions, and observational constraints. Despite differences in assumptions and outcomes, all models aim to describe matter at supra-nuclear densities and must satisfy constraints from massive neutron stars (\(\geq 2M_{\odot}\)), NICER radius measurements, and tidal deformability data from events like GW170817.
	
	Most models agree that nucleonic matter dominates at lower densities, while exotic components such as hyperons, meson condensates, or quark matter may appear at higher densities. Hadronic models, especially relativistic mean-field (RMF) models, describe stars composed of interacting nucleons and typically yield stiff EoSs, though the inclusion of hyperons or mesons softens the EoS. Hybrid models introduce a transition to quark matter, with crossover models like QHC21 offering smooth transitions, while first-order phase transition models may lead to mass twins. Pure quark matter models, including NJL, FRG, and V-QCD frameworks, suggest deconfined quark cores or entirely quark-based stars, but often require strong repulsive interactions to remain consistent with \(2M_{\odot}\) observations.
	
	While no single model is definitively confirmed, overly soft EoSs are largely disfavored, whereas stiff EoSs and hybrid models with phase transitions remain promising. The continued integration of astrophysical observations, nuclear experiments, and QCD theory is essential for advancing our understanding of dense matter and potential exotic phases inside neutron stars.

	\bibliographystyle{plain}

\end{document}